\documentclass[%
 reprint,
 amsmath,amssymb,
 prl,
]{revtex4-1}

\usepackage{graphicx}
\usepackage{dcolumn}
\usepackage{bm}
\usepackage{float}
\usepackage{xcolor}
\usepackage{gensymb}
\usepackage{amsmath}
\usepackage{hyperref}

\begin{document}

\preprint{APS/123-QED}

\title{Sub-second production of a quantum degenerate gas}

\author{Gregory A. Phelps}
\author{Anne H\'{e}bert}
\author{Aaron Krahn}
\author{Susannah Dickerson}
\author{Furkan \"{O}zt\"{u}rk}
\author{Sepehr Ebadi}
\author{Lin Su}

\author{Markus Greiner}
\email{greiner@physics.harvard.edu}
\affiliation{%
Department of Physics, Harvard University, Cambridge, Massachusetts 02138, USA
}%

\date{\today}

\begin{abstract}
Realizing faster experimental cycle times is important for the future of quantum simulation. The cycle time determines how often the many-body wave-function can be sampled, defining the rate at which information is extracted from the quantum simulation.  We demonstrate a system which can produce a Bose-Einstein condensate of $8 \times 10^4$ $^{168}\text{Er}$ atoms with approximately 85\% condensate fraction in 800 ms and a degenerate Fermi gas of $^{167}\text{Er}$ in 4 seconds, which are unprecedented times compared to many existing quantum gas experiments. This is accomplished by several novel cooling techniques and a tunable dipole trap. The methods used here for accelerating the production of quantum degenerate gases should be applicable to a variety of atomic species and are promising for expanding the capabilities of quantum simulation.
\end{abstract}

\maketitle

The realization of quantum systems which push beyond the limits of classical simulation is at the forefront of modern physics research \cite{Arute2019}. Ultracold atoms are a particularly advanced platform for quantum simulation, as they provide excellent many-body coherence and, through quantum gas microscopy, high fidelity single atom control \cite{Bakr2010, Sherson2010,Haller2015, Parsons2015,Miranda2015,Yamamoto2016, Omran2015, Cheuk2015a, Edge2015, Greif2016, Kondov2017, Lukin2019}. Fermionic atoms even allow the native realization of fermionic physics models, such as the Hubbard model, which would require a large overhead in qubit based quantum simulations \cite{Barends2015, Steudtner2019}. While the particle number in ultracold atom systems can be very large, a major limitation of the traditional approach of evaporative cooling is the slow cycle time of the experiments, typically on order of 10 seconds to a minute. This is particularly a hindrance for state-of-the-art experiments that directly sample from the many-body wave-function and extract high order correlation functions \cite{Schweigler2017,Rispoli2019}, detect patterns \cite{ Boll2016a, Mazurenko2016, Chiu2019} or apply machine learning \cite{Bohrdt2019}. 

Significant work has gone into reducing the cycle times of experiments either by optimizing evaporation \cite{Kinoshita2005, Roy2016} or by all-optical cooling techniques  \cite{Stellmer2013, Solano2019, Hu2017,Urvoy2019}.  Optimizing evaporation through dynamical trap shaping speeds up cycle time (1.6--3.3 s), builds on well-established techniques, and has the advantage of being relatively simple to implement in existing systems, but requires significant atom loss to remove energy from the system.  On the other hand, all-optical cooling has resulted in extremely fast cooling (0.3--2 s) to quantum degeneracy with low atom loss, but have lower condensate fractions of 7--40\% and low atom number in the case of degenerate Raman sideband cooling in a lattice.  Further evaporative cooling is likely necessary in these schemes to reach condensate fractions useful for most experiments.  A hybrid approach works with both optical and evaporative cooling, which has resulted in relatively fast cooling (2 s) to quantum degeneracy \cite{Stellmer2012} while still retaining large atom number.

\begin{figure}[h!]
    \centering
    \includegraphics[width=8.6cm]{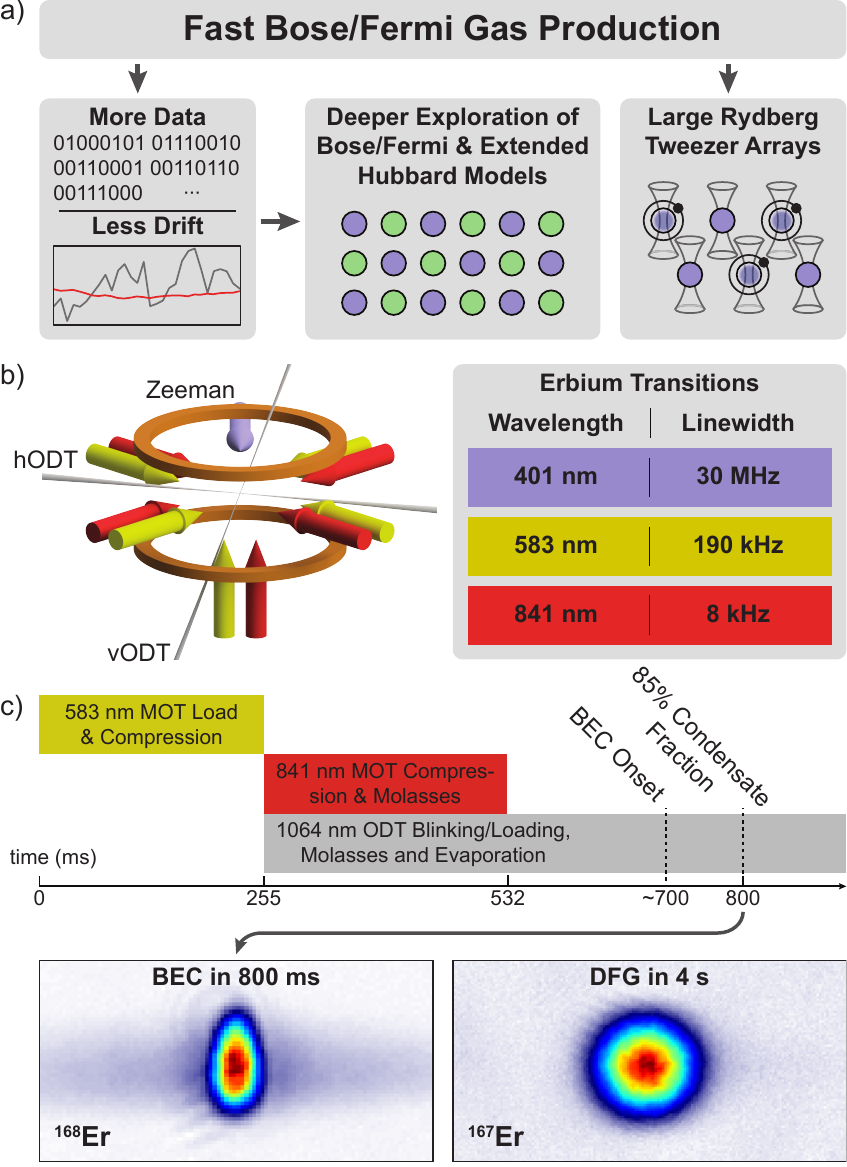}
    \caption{Illustration of the benefits of a fast repetition rate, as well as an illustration of setup and sequence of the experiment.  (a) Fast BEC/Fermi gas production allows for more data and less drift in the data, which will allow the Bose-Hubbard, Fermi-Hubbard, and Extended Hubbard models to be explored in ways previously inaccessible to quantum gas microscopes. We also propose using this method to generate large arrays of tweezers without the need for resorting or cooling. (b) An illustration of the setup including magnetic coils, 583 nm and 841 nm MOT beams, horizontal (hODT) and vertical (vODT) optical dipole trap beams, and the Zeeman slowing beam.  The transitions wavelengths are given including respective transition linewidths. (c) An illustration of the experimental sequence used to produce a Bose-Einstein condensate (BEC) in 800 ms.  A similar sequence produces a degenerate Fermi gas (DFG) in 4 seconds. }
    \label{fig:figure1}
\end{figure}

In this letter we present a method to rapidly create ultracold quantum gases with a large atom number. Specifically, we generate dipolar Bose-Einstein condensates (BEC) of $8\times 10^4$ erbium atoms with 85\% condensate fraction within 800 ms and dipolar degenerate Fermi gases (DFG) within 4 seconds (see Figure \ref{fig:figure1}). This corresponds to a speedup over typical experiments of one to two orders of magnitude, and will enable experiments with itinerant quantum gases with cycle times in the Hertz range. Furthermore, the cycle time is comparable to recent optical tweezer Rydberg systems, albeit with much larger atom numbers \cite{Barredo2016,Endres2016,Cooper2018a, Norcia2018}. The speed of our method also opens the possibility of loading much larger Rydberg tweezer arrays directly from a Mott insulator, without the need for tweezer sorting or Raman sideband cooling \cite{Kaufman2012, Endres2016,Cooper2018a}.  Our method is based on novel narrow-line laser cooling techniques and evaporative cooling in a dipole trap with tunable geometry. Furthermore, our method is applicable to both the bosonic and fermionic isotopes of erbium.

Our experimental setup uses three optical transitions in erbium for slowing and cooling the atoms and a tunable dipole trap operating at 1064 nm (see Figure \ref{fig:figure1}b.  A BEC is generated by initially loading $^{168}\text{Er}$ into a 5 beam narrow-line magneto-optical trap (MOT) using the 583 nm transition \cite{BAN2005, Berglund2008,Frisch2012, Ilzhofer2017}, which has a linewidth of $\Gamma_{\text{583}}  \approx 2\pi \times 190$ kHz.  This MOT naturally polarizes the atoms into the high-field seeking stretched state.  The initial capture of erbium atoms in the MOT is facilitated by an effusion oven (operating at 1100 \degree C) followed by a 2D molasses and a spin-flip Zeeman slower, both operating on the 401 nm transition - the 401 nm transition has a linewidth of $\Gamma_{\text{401}}  \approx 2\pi \times 30$  MHz. After loading of the 583 nm MOT, the atoms are then handed-off to a red-detuned 841 nm MOT ($\Gamma_{\text{841}}  \approx 2\pi \times 8$ kHz) where the atoms are further cooled and simultaneously loaded into an optical dipole trap (ODT).  After a second cooling step with the 841 nm light, the atoms undergo plain evaporation to quantum degeneracy.

\begin{figure}[h!]
    \centering
    \includegraphics[width=8.6cm]{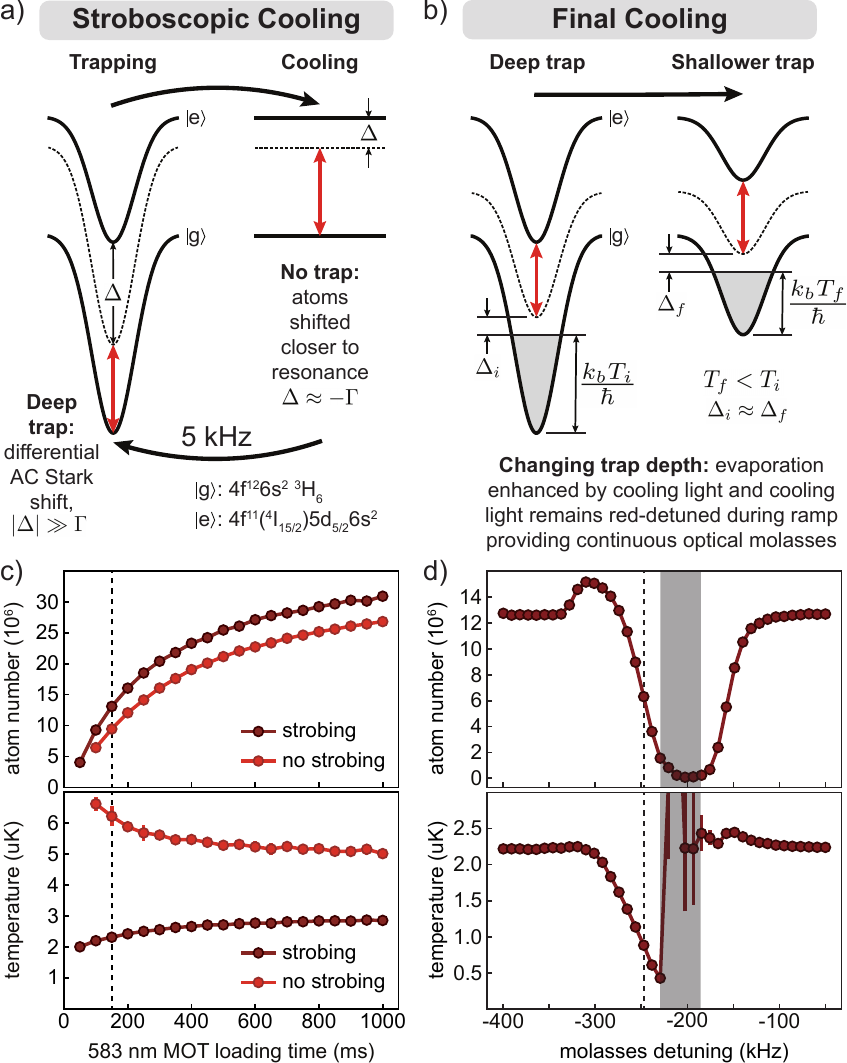}
    \caption{Narrow-line cooling methods.  (a) Illustration of the stroboscopic cooling technique used to overcome a differential AC Stark shift.  (b) Illustration of the final cooling step in which evaporation and optical molasses occur simultaneously.  841 nm light is drawn from excited state to show frequency splitting relative to atoms in the electronic ground state and is fixed in absolute frequency during the ramp. (c) Atom number and temperature during stroboscopic loading of the ODT as a function of 583 nm MOT load time. Atom number is increased and temperature is decreased when compared to loading without strobing the ODT light. (d) Atom number and temperature in horizontal ODT versus 841 nm molasses detuning.  Greyed region represents the range of resonance conditions satisfied during the ramp down of the trap depth. Detuning here is relative to the bare resonance without Stark shifts or Zeeman shifts.  Note the dashed lines in (c) and (d) represent the MOT loading time and molasses  detunings used in the experiment to produce a fast BEC.}
    \label{fig:figure2}
\end{figure}

Reducing experimental cycle times often requires every detail of the experiment to be operating at peak capabilities.  One significant limitation in cycle time is MOT loading time.  In this experiment we pushed to increase the loading rate of the 583 nm MOT \cite{*[{See supplemental material below for additional experimental details, ramps, and trimodal fitting}] [{}] Supplementary2020}.  For generating a $^{168}\text{Er}$ BEC, we load the 583 nm MOT for 150 ms at a rate of $250 \times 10^6 \text{ atoms/s}$.   We achieve this loading rate by careful choice of beam parameters for the 2D molasses and a broadband MOT \cite{Katori1999, Supplementary2020}.  The 2D molasses has the largest contribution to loading rate increase and has afforded us a 30 times increase in loading rate. The broadband MOT technique also increases the loading rate an additional 2-3.  Note there also exists a core-shell technique using an even broader transition to help slow atoms \cite{ Lee2015}, which achieves impressive results and can be implemented using the 401 nm transition in erbium. We rely on a broadband MOT, since it limits additional optics on the experiment table and is easily implemented in existing systems.  

We further decrease the cycle time by implementing a second stage, narrower-line red-detuned MOT operating on the 841 nm transition.   This MOT serves to increase the initial PSD loaded into the ODT.  Similar narrow-line MOTs have been implemented using the Strontium $^1 {S}_0$--$^3 {P}_1$ transition at 689 nm and can lead to an initial PSD of $\mathcal{D} = 0.1$ in an ODT \cite{Stellmer2012}.  Additionally, a blue-detuned MOT has been previously implemented using the 841 nm transition in erbium \cite{Berglund2008}.   We were able to operate in the red-detuned regime, since we load the 841 nm MOT from a 583 nm MOT, which already cools the atoms to nearly 10 uK.  We achieve nearly a 100\% hand-off efficiency by simply turning the 583 nm MOT off and the 841 nm MOT on at the same time.  Interestingly, when quantifying the narrowness of a  MOT by the ratio of the maximum optical force to the downwards force of gravity, $\eta = \frac{\hbar k \Gamma}{2 m g}$, this MOT is the narrowest in operation with $\eta \approx 7$. The 841 nm MOT has allowed us to reach temperatures of nearly 400 nK, which is far below the capability of the 583 nm MOT and what was shown by Berglund, et al \cite{Berglund2008}.  In addition, we estimate the peak PSD achieved in the 841 nm MOT to be $\mathcal{D} \approx 0.05$, which is two orders of magnitude higher than $\mathcal{D} \approx 3\times10^{-4}$ achieved in the 583 nm MOT.    

\begin{figure}[h!]
    \centering
    \includegraphics[width=8.6cm]{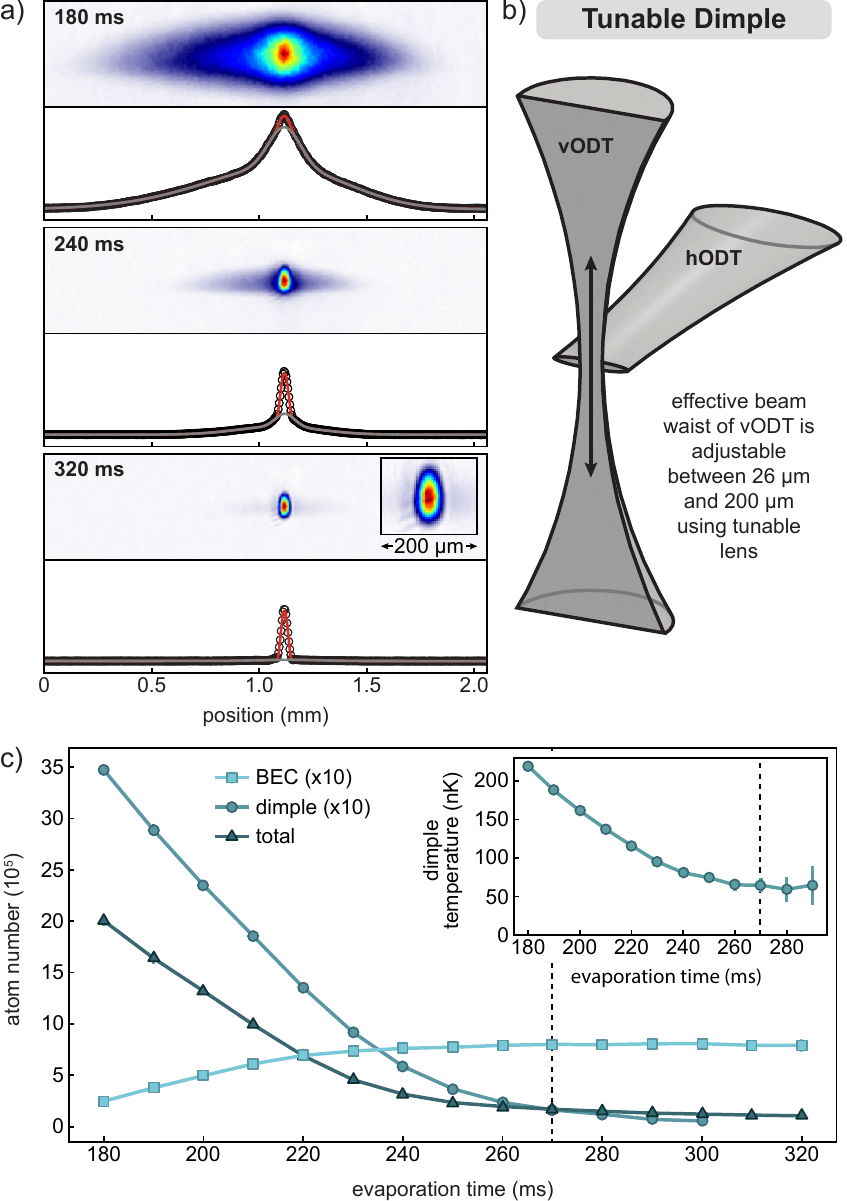}
    \caption{Evaporation of $^{168}\text{Er}$ to degeneracy. (a) Three averaged time-of-flight (TOF) images (taken at 22.3 ms TOF) of the evaporation starting just after BEC onset and ending 140 ms later at a nearly pure BEC. Below each TOF image is the integrated density distribution (circles) showing the fitted (see Supplementary \cite{Supplementary2020} for description of fitting functions) BEC (red) and excluding the BEC (grey).  With careful inspection atoms can be seen falling out of the ODT at the 240 ms evaporation time. (b) An illustration of the dipole trap configuration showing the direction in which the waist of the vODT is moved to effectively change the dimple size.  (c) The atom number and temperature during the last part of evaporation.  The BEC and dimple atom numbers are multiplied by 10 for better visibility. Dashed line corresponds to 800 ms total cycle time.}
    \label{fig:figure3}
\end{figure}

To further speed up the cycle time, atoms are simultaneously loaded into a 1064 nm horizontal optical dipole trap (hODT) during 841 nm MOT compression as shown in Figure \ref{fig:figure1}b,c.  The hODT has beam waists of approximately $19 \text{ }\mu\text{m} \times 550 \text{ }\mu\text{m}$ with the smaller waist oriented along gravity \cite{Supplementary2020}. Loading a large PSD into the hODT initially proved challenging due to a differential AC Stark shift causing the excited state polarizability to be approximately 60 percent of the ground state.  To overcome this limitation we implemented a technique which strobes the hODT light intensity while the MOT loads into the hODT \cite{Shih2013, Hutzler2016}.  We strobe with the light 160 $\mu$s on and 40 $\mu$s off, which allows simultaneous trapping and cooling \cite{Supplementary2020}. This technique partially overcomes the limitations introduced by the differential AC Stark shift by providing a short period of cooling as shown in Figure \ref{fig:figure2}a.  Strobing the hODT requires more power than would normally be necessary for the same trap depth; however, the atoms can be loaded at a lower temperature into the hODT as is shown in Figure \ref{fig:figure2}c.  In this experiment the average power during loading is approximately 10 W, which corresponds to 18 $\mu$K.  This yields a temperature to trap depth ratio of approximately 9 - a good starting point for efficient evaporation.  Strobing has proven to be a powerful tool which is easily implemented in existing experiments and may provide large additional PSD increases (an order of magnitude in our case) for modest increases in optical power.

Another bottleneck for most quantum gas experiments is evaporation to a degenerate gas.  Efficient, as well as fast evaporation, requires maintaining a high ratio of elastic to inelastic collisions.  This can be difficult with a fixed trap geometry, since the only controllable parameter is trap depth.  Several techniques to control trap geometry has enabled faster evaporation to a degenerate gas \cite{Roy2016, Kinoshita2005}. We implemented a 1064 nm tunable dimple using a small, nearly symmetric, vertically-oriented optical dipole trap (vODT), aligned to the middle of the hODT. The minimum waist size of the vODT is approximately 26 $\mu$m, which was chosen to allow the large trapping frequencies necessary for fast evaporation. The position of the vODT focus can be tuned (see Figure \ref{fig:figure3}b) with a tunable lens, which provides tunability of the effective waist size up to 200 $\mu$m  \cite{Supplementary2020}.   

After the hODT is loaded, a magnetic field is applied to maintain spin purity of the cloud. The vODT is simultaneously turned on at 1.4 W and fixed in waist size at approximately 60 $\mu$m, which corresponds to a dimple depth of 9 $\mu$K.  We found that we could further increase the PSD and help to load the dimple by ramping down the hODT from 10 W to 7.25 W in 200 ms and applying an 841 nm optical molasses at the same time.  In this cooling step the light is slightly red-detuned from atoms in the hODT.  The atoms in the dimple are shifted far enough from resonance that they do not scatter significant light from the molasses.  The mechanism of cooling is illustrated in Figure \ref{fig:figure2}b.  As the trap reduces in depth the atoms undergo the standard optical evaporation, but the speed is enhanced by the molasses light.  In Figure \ref{fig:figure2}d the optical molasses frequency is changed and atom number and temperature is measured in the hODT.  In this figure the vODT is off to show the cooling effect of the atoms on the hODT.  A PSD gain of nearly an order of magnitude can be achieved at an efficiency of nearly 2.5.  While this is not as efficient as is possible without the molasses, it is significantly faster.  The addition of the vODT serves to further enhance the PSD and atom number.  At the end of the 200 ms ramp the PSD in the dimple reaches $\mathcal{D} \approx 0.07$, the atom number in the dimple is $1\times 10^6$ and the total atom number in the hODT and dimple is $5\times 10^6$.  The efficiency of this step with the vODT present is approximately 3.0.  In principle, this step is limited in efficiency due to the differential AC Stark shift which forces some of the atoms to see blue-detuned light and are heated.  

The final cooling step uses evaporative cooling and occurs over a period 270 ms.  The hODT is ramped in power from 7.25 W to 1.2 W \cite{Supplementary2020}.   At the same time the dimple size is quickly reduced to approximately 30 $\mu$m waist and the power is reduced.  The evaporation is accelerated by the favorable scattering cross section of approximately $200 a_0$ found in $^{168} \text{Er}$ \cite{Aikawa2012}.  BEC onset occurs approximately 180 ms into the evaporation sequence as is shown in Figure \ref{fig:figure3}.  At this point $2\times 10^6$ atoms remain in the hODT and dimple.  The atom number in the BEC quickly saturates after onset and reaches a condensate fraction of 85\% after 270 ms (800 ms cycle time).  After an additional 20 ms of evaporation the condensate fraction reaches approximately 92\% and the thermal fraction can no longer be reliably identified.  Additional evaporation removes thermal atoms from the hODT and does not visibly change the condensate in the dimple.  It is worth noting that we are able to produce pure BECs of $2\times 10^4$ atoms in under 700 ms total cycle time by shortening MOT loading time and evaporation.

\begin{figure}[h!]
    \centering
    \includegraphics[width=8.6cm]{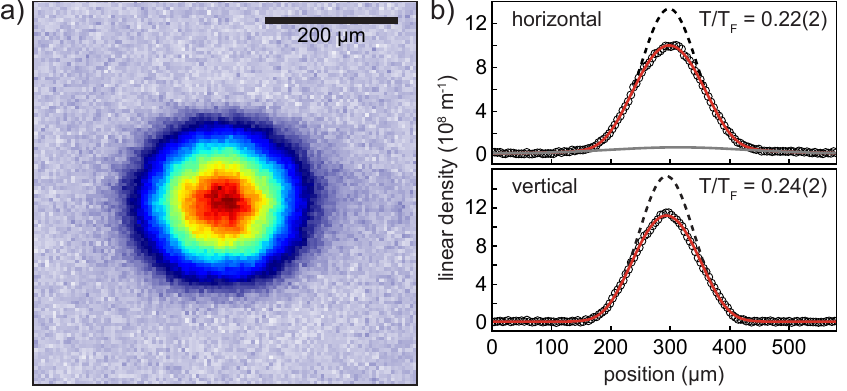}
    \caption{Evaporation of  $^{167}\text{Er}$ to a degenerate Fermi gas. (a) A single absorption image of $N = 1.3\times 10^5$ atoms after a time-of-flight (TOF) expansion of $t_{\text{TOF}} = 22 \text{ ms}$. (b) The corresponding integrated density distribution along the horizontal and vertical axes at $T/T_F = 0.22(2)$ and $T/T_F = 0.24(2)$, respectively. The integrated distributions (circles) are fitted with a poly-logarithmic plus Gaussian function (solid red line); the Gaussian part of this fit is shown in grey and it corresponds to the thermal atoms in the sheet along the horizontal axis.  An additional fit with only a Gaussian function (dashed line) to the tails of the distribution is shown in order to emphasize the deviation from a classical gas. }
    \label{fig:figure4}
\end{figure}

One of the more promising aspects of this trap configuration, when applied to erbium, is the ability to evaporate multiple isotopes. As is shown by Aikawa, et al.  \cite{Aikawa2014a}, the evaporation of the fermionic isotope $^{167}\text{Er}$ proved more challenging and required the use of a 1570 nm laser instead of the 1064 nm laser used for the bosonic isotopes.  Since the fermionic isotope is loaded into the dipole trap at relatively low temperatures, the need for the 1570 nm laser has been circumvented and we have produced degenerate Fermi gases of $1.5\times 10^5$ atoms with $T/T_F \approx 0.25$ with a total cycle time of approximately 4 s. While this is not as fast as the BEC generation, due in part to the reduced effective scattering length compared to $^{168}\text{Er}$, it is a substantial improvement over existing dipolar Fermi gas production \cite{Desalvo2010,Lu2012, Stellmer2012, Aikawa2014a}. The dipolar fermionic isotopes may prove to have a significant advantage over other species due to the non-zero scattering cross-section at zero temperature \cite{Aikawa2014a}. In addition to DFGs, larger pure dipolar $^{168}\text{Er}$ BECs of up to $10^6$ atoms can be produced by increasing the MOT loading time and increasing the evaporation time to a total cycle time of 5 s.  The size of these BECs are limited by the large atom number transfer efficiency from the 841 nm MOT to the hODT as well as the cooling step.  We believe that all isotopes of erbium can be evaporated by appropriate choice of ramps, which will make this system useful for studying even the least abundant isotope, $^{162}\text{Er}$.

In conclusion, we have pushed the limits of several standard techniques to produce ultracold atoms and developed several new techniques to aide in the production of degenerate quantum gases.   Specifically, we have implemented an ultra-narrow red-detuned MOT using the 841 nm transition in erbium, which is operating near the limit of support against gravity, developed several cooling techniques to help cool atoms in an ODT in the presence of a differential AC stark shift, and implemented a tunable dipole trap using a tunable lens.  All of this work has culminated in the fastest large BEC production time to date. This marks a significant improvement over existing systems. It should possible to further reduce the cycle time for every isotope with this setup with additional work on finding a magic ODT wavelength for the 841 nm transition, improving MOT loading, implementing a tunable hODT to overcome lower scattering cross-sections of different isotopes, and using machine-learning to optimize evaporation ramps \cite{Barker2020}.

We would like to thank Francesca Ferlaino and the ERBIUM and Er-Dy teams in Innsbruck for their collaboration. We thank Alex Lukin for helpful discussions, as well as Sambuddha Chattopadhyay, Adam Ehrenberg, Maximilian Sohmen from Innsbruck, and Emily Tiberi for their work on the experiment.  We acknowledge support from NSF (CUA, GRFP).  AFOSR (MURI), ARO (MURI, DURIP), ONR (Vannevar Bush Faculty Fellowship), DARPA, and the Gordon and Betty Moore foundation EPiQS initiative. 

A.H. and A.K. contributed equally to this work.


%

\clearpage
\newpage

\setcounter{page}{1}
\setcounter{figure}{0}
\renewcommand{\thepage}{S\arabic{page}} 
\renewcommand{\theequation}{S\arabic{equation}}
\renewcommand{\thefigure}{S\arabic{figure}}
\renewcommand{\bibnumfmt}[1]{[S#1]}

\onecolumngrid

\begin{center} 
	\begin{large} \textbf{Supplemental material: Sub-second production of a quantum degenerate gas} \newline \end{large}
\end{center}

\twocolumngrid

\section{Additional Experimental Details}

\begin{figure*}[h!]
    \centering
    \includegraphics[width=17.2cm]{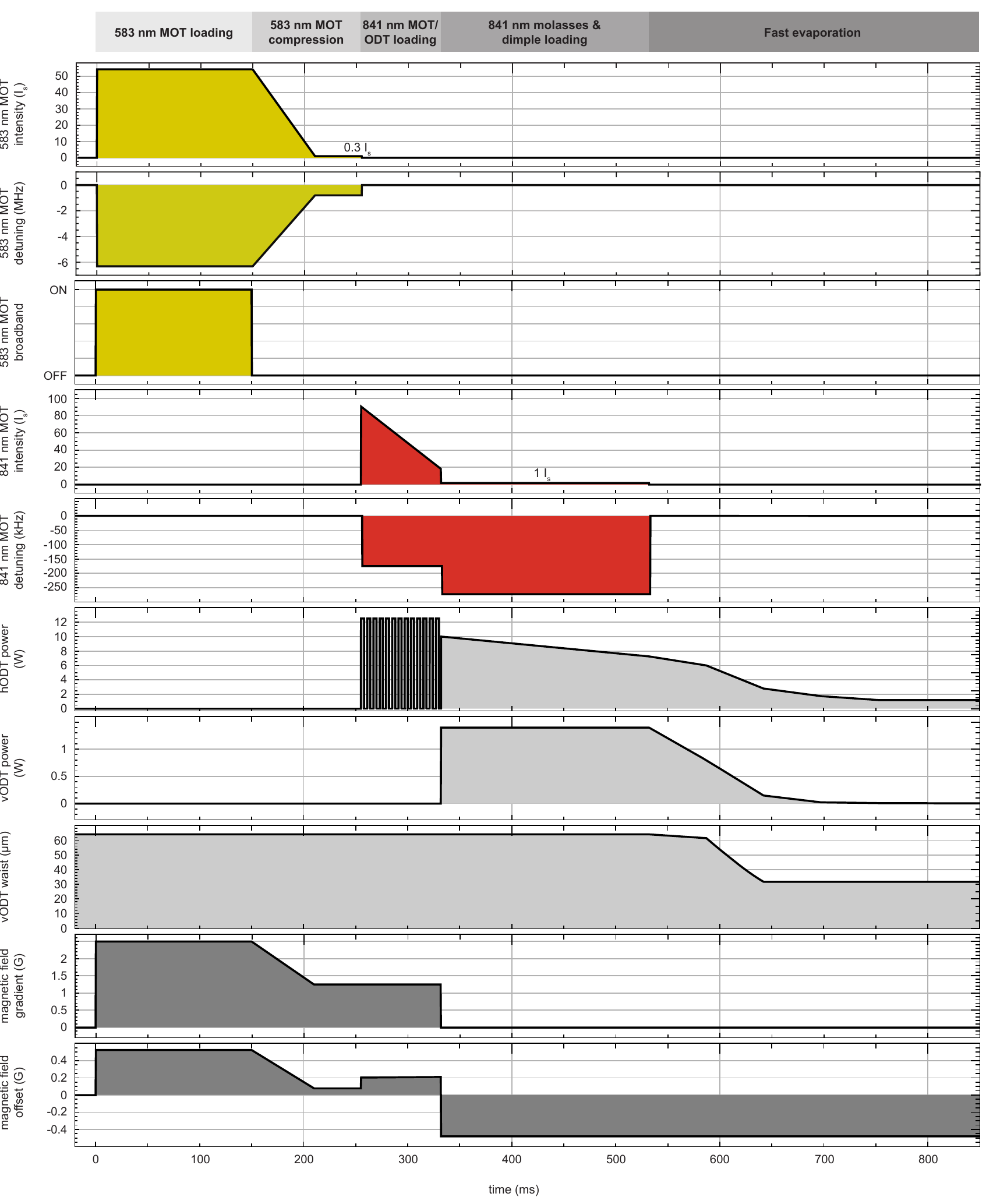}
    \caption{Plots showing experimental sequence used to achieve fast BEC production.  A short description of each part of the sequence can be found at the top of the figure.  Each plot is color coded and grouped to illustrate the effect of each operation.  Operations done to 583 nm MOT are in yellow, 841 nm are in red, ODT are in light grey, and magnetic fields are in dark grey. Note the strobing depicted in the hODT power plot is only for visualization - the frequency and duty cycle are not accurate.}
    \label{fig:sequence_fig}
\end{figure*}

A figure of the full experimental sequence for producing a BEC in 800 ms can be found in Figure \ref{fig:sequence_fig}.  The figure includes the light intensity, frequency, dimple size, and magnetic fields used in the experiment.

The flux of Erbium atoms emanate from an oven (CreaTec DFC-40-10-WK-2B-SHP) operating at 1100 \degree C for the effusion cell and 1200 \degree C for the hot lips.  A 2D molasses, directly after the oven aperture, operating on the 401 nm transition, is used to increase the atom-flux into the MOT chamber. The beam waists are approximately $3\text{mm} \times 15 \text{mm}$, with the large waist oriented along the atom-beam line to increase interaction time.  Each beam has a peak intensity of approximately $2 I_{s}^{\text{401}}$, where $I_s^{\text{401}} \approx 60 \text{ mW}/\text{cm}^2$ is the saturation intensity of the 401 nm transition.  The 2D molasses increases the flux into the science chamber (and subsequently the loading rate into the magneto-optical trap) by a factor of 30. The atoms are slowed from the initial velocity of approximately $400 \text{ m}/\text{s}$ by a spin-flip Zeeman slower also operating on the 401 nm transition. 

The broadband MOT technique modulates the frequency of the light by modulating the frequency of the acousto-optic modulator (AOM) supplying the light to the MOT.  This is achieved by mixing an 80 MHz carrier with a 30 MHz signal from a signal generator to generate the required 110 MHz signal for the AOM.  The FM port of the signal generator is then modulated by an external function generator with a triangle wave.  The amplitude of the triangle wave is controlled by our experimental control software.  Typically, broadband MOT implementations use an external electro-optic modulator (EOM) for modulation, but this shows the same results can be achieved with a simpler upgrade, using inexpensive equipment and not reworking any existing optical setups. 

The 583 nm MOT during loading operates at approximately $I \approx 50 I_{\text{s}}^{583}$, an average detuning of -6.3 MHz, and a gradient of -2.5 G/cm. Due to the narrow-line nature of this MOT, the MOT operates nearly the same if the downwards MOT beam is removed; therefore, to free up optical access, we operate in this 5 beam configuration.  The compression of the 583 nm MOT occurs after just 150 ms of loading. The broadband scanning is turned off and a linear ramp takes the MOT to its final compressed state with $I \approx 0.3 I_{\text{s}}^{583}$, a detuning of -800 kHz and a gradient of 1.25 G/cm.  The gradient is reduced since atoms are lost more quickly from the 841 nm MOT at higher gradients, due to the high field seeking state that atoms are polarized into.

The hand-off to the 841 nm MOT is simple and just requires matching the frequency of the 841 nm MOT to the position of the 583 nm MOT and jumping an offset field slightly. An offset field is applied to help control the position of the MOT during the ODT loading.  The 583 nm light is turned off during hand-off and the 841 nm light is turned on to approximately $I = 100 I_{\text{s}}^{841}$.  The 583 nm MOT atoms are allowed to fall slightly into the 841 nm MOT, which allows the 841 nm MOT to catch all atoms from the 583 nm MOT.  This hand-off is consistent with 100\% efficient when natural losses from the compressed MOTs are accounted for.  

The 1064 nm ODTs are derived from an ALS 1064 nm fiber laser with the ALS external seeder option.  The light for the hODT is delivered through a 2 m photonic crystal fiber (NKT LMA-PM-15), which is used to maintain alignment of the hODT to a higher degree and avoid any drifts from the laser.  It was important to use a 2 m fiber here since we are putting up to 14 W through the fiber. We have seen stimulated Brillouin scattering (SBS) when doing the same with a 5m fiber.  The light going to the vODT is delivered through a Thorlabs polarization maintaining fiber, but the power is limited to 1.5 W.  We have not seen any degradation of this fiber nor SBS.  The two ODT beams are frequency shifted from each other by AOMs, which also function to control intensity in a closed-loop feedback.  Strobing the hODT is achieved by externally modulating the RF to the AOM. This does effect the feedback loop, but we circumvent this by using a low pass filter with a 3 dB point of approximately 200 Hz. 

The hODT has beam waists of approximately $19 \text{ }\mu\text{m} \times 550 \text{ }\mu\text{m}$.  The smaller waist is oriented along gravity and provides a trapping frequency in the vertical direction of nearly 600 Hz.  The larger waist was chosen to have a better mode-overlap with the 841 nm MOT, whereas the smaller waist was chosen to allow larger trapping frequencies for evaporation.  

The vODT has a minimum waist of approximately 26 $\mu$m and is tunable up to 200 $\mu$m using a tunable lens (Optotune EL-16-40-TC-20D with IR coating).  The beam is angled approximately 20 degrees relative to vertical due to constraints from other optics.  The tunable lens is placed in the Fourier plane of a focusing lens which allows the beam waist to be moved without changing the minimum waist of the beam.  We achieve fine control of the focus position by using a short focal length lens after the tunable lens and a custom current driver, which provides up to 40 mA of current to the tunable lens with a 0--10 V input signal.  

During the 841 nm compression phase we turn on the hODT and strobe the light from it at 5 kHz with an 80\% duty cycle (80\% on, 20\% off).  This helps to increase loading into the ODT and reduces the temperature in the ODT significantly.  It is important to choose the strobing frequency, $f_{\text{strobe}}$, to be much larger than the trapping frequency, $f_{\text{strobe}} \gg f_{\text{trap}}$, but small enough to allow for scattering to occur during an off period of the trap. This allows the atoms to see a time-averaged potential and also be cooled.  For certain situations it is important to allow time for the MOT light to be off just after a cooling pulse, but before a trapping pulse.  Specifically, the light should be off for at least one decay time to prevent the atom from being trapped in the excited state, otherwise Sisyphus heating may be present.  This is not a concern for our experiment as the trap frequencies are sufficiently low (maximum of 600 Hz) that the atoms do not move much in the excited state before decaying back down. The compression phase consists of ramping down the intensity of the light from $100 I_{\text{s}}^{841}$ to $20 I_{\text{s}}^{841}$.  Because of the narrow-line nature of the 841 nm transition, we only have to move the position of the MOT slightly (with magnetic fields) during compression.  Doing so guarantees good overlap with the ODT throughout compression.  

After the loading stage, we apply a homogeneous magnetic field of approximately 0.5 G to the cloud to preserve spin. We change the frequency of the 841 nm laser to compensate for the AC stark shift from the ODT and the Zeeman shift from the magnetic field.  The intensity of the light is also reduced to $I \approx 1 I_{\text{s}}^{841}$.  We then proceed to leave all values fixed, except we ramp the power of the hODT from 10 W to 7.25 W, which serves to cool the cloud through molasses and evaporation. All of this occurs over 200 ms.

After this stage we enter a plain evaporation stage in which we ramp down the intensity of both the hODT and vODT, as well as reduce the effective waist size of the vODT.  The waist is reduced rather quickly in the beginning and then held at a fixed value for the remainder of evaporation. The large initial size of the vODT creates a good overlap with the atoms in the hODT and allows atoms to quickly and efficiently collect in the dimple.  The dimple size is then reduced, which compresses the sample and and increases the collision rate.  After the dimple is fixed in size, both the hODT and vODT ramp down in a nearly exponential fashion. Approximately 50 ms after the dimple is fixed in size we have onset of BEC.

\newpage

The state of the system is then measured using standard absorption imaging after 22.3 ms time of flight. The images of the cloud are extracted using either a low resolution imaging system with a Point Grey Grasshopper 3 camera or a higher resolution imaging system using an ANDOR iXON 897.  We use fast-kinetics mode on the iXON, since it offers lower noise and faster effective frame rates. This reduces fringing effects from air currents, as well as effects from light intensity changes. 

\section{Thermal and BEC fitting}

Fitting of the 1D integrated profile (integrated in the vertical direction) requires a trimodal fit accounting for the BEC, dimple and sheet separately. Temperature, atom number, and condensate fraction are extracted from these fits.  The BEC component is fit by the standard integrated Thomas-Fermi distribution 

\begin{equation}
    \phi_{\text{BEC}}(x) =
    \begin{cases}
    \frac{15 N_{\text{BEC}}}{16 w}\left(1-\left(\frac{x-x_0}{w}\right)^2\right)^2,\; |x-x_0| < w  \\
    0, \text{ otherwise,}   \\
    \end{cases}
\end{equation}

\noindent where $N_{\text{BEC}}$ is the atom number in the BEC, $x_0$ is the offset, and $w$ is the width.  The dimple fit uses the standard Gaussian fit for the thermal componenet, which uses the same offset as the BEC.  The sheet fitting is complicated by aberration in the hODT, which causes non-gaussian density distribution. To compensate this, a skew-normal distribution is used for fitting the sheet  

\begin{equation}
    \phi_{\text{sheet}}(x) = \frac{N_s}{\sqrt{2 \pi} w_s} e^{-\frac{(x-x_s)^2}{2 w_s^2}} \left(1+\text{erf}\left(\alpha \frac{x-x_s}{\sqrt{2} w_s}\right)\right)
\end{equation}

\noindent where $N_s$ is number of atoms in the sheet, $x_s$ is the offset of the sheet, $w_s$ is the width of the sheet, and $\alpha$ is the shape parameter. This distribution was chosen because it yielded the best fit results at both higher and lower temperature.

\clearpage

\end{document}